\documentclass[preprint]{sigplanconf}
\usepackage[color]{changebar}
\usepackage{enumitem}
\usepackage[lighttt]{lmodern}
\usepackage{amsmath}
\usepackage{amssymb}
\usepackage{amsfonts}
\usepackage{graphicx}

\usepackage{listings}
\usepackage{lmodern}
\ttfamily
\DeclareFontShape{OT1}{lmtt}{m}{it}
     {<->sub*lmtt/m/sl}{}
\lstdefinestyle{default}{language=Anglican, basicstyle=\ttfamily, columns=flexible, showstringspaces=false}

\usepackage{hyperref}
\usepackage{url}

\usepackage{listings}
\usepackage{color}
\usepackage{xcolor}
\definecolor{blue}{rgb}{0,0.3,0.7}
\definecolor{red}{rgb}{0.60,0.0,0.0}
\definecolor{purple}{rgb}{0.5,0,0.7}
\definecolor{cyan}{rgb}{0.0,0.6,0.5}
\definecolor{gray}{rgb}{0.4,0.4,0.4}

\lstdefinelanguage{scheme}
{sensitive, %
 alsoletter={:,-,+,*,?,/,!,>,<}, %
 morecomment=[l];, %
}[comments]

\lstdefinelanguage{anglican}%
{
 morekeywords=[1]{},
 morekeywords=[2]{%
   def, def-, defn, defn-, defmacro, defmulti, defmethod, %
   defstruct, defonce, declare, definline, definterface, %
   defprotocol, defrecord, defstruct, deftype, defproject, ns, %
 }, %
 morekeywords=[3]{->, ->>, .., amap, and, areduce, as->, assert, binding, %
   bound-fn, case, comment, cond, cond->, cond->>, condp, declare, definline, %
   definterface, defmacro, defmethod, defmulti, defn, defn-, defonce, %
   defprotocol, defrecord, defstruct, deftype, delay, doseq, dosync, dotimes, %
   doto, extend-protocol, extend-type, fn, for, future, gen-class, %
   gen-interface, if, if-let, if-not, if-some, import, io!, lazy-cat, lazy-seq, let, %
   letfn, locking, loop, memfn, ns, or, proxy, proxy-super, pvalues, %
   recur, refer-clojure, reify, some->, some->>, sync, time, when, when-first, %
   when-let, when-not, when-some, while, with-bindings, with-in-str, %
   with-loading-context, with-local-vars, with-open, with-out-str, %
   with-precision, with-redefs}, %
  morekeywords=[4]{*, *', +, +', -, -', ->ArrayChunk, ->Vec, ->VecNode, %
    ->VecSeq, -cache-protocol-fn, -reset-methods, /, <, <=, =, ==, >, >=, %
    accessor, aclone, add-classpath, add-watch, agent, agent-error, %
    agent-errors, aget, alength, alias, all-ns, alter, alter-meta!, %
    alter-var-root, ancestors, apply, array-map, aset, aset-boolean, aset-byte, %
    aset-char, aset-double, aset-float, aset-int, aset-long, aset-short, assoc, %
    assoc!, assoc-in, associative?, atom, await, await-for, await1, bases, bean, %
    bigdec, bigint, biginteger, bit-and, bit-and-not, bit-clear, bit-flip, %
    bit-not, bit-or, bit-set, bit-shift-left, bit-shift-right, bit-test, %
    bit-xor, boolean, boolean-array, booleans, bound-fn*, bound?, butlast, byte, %
    byte-array, bytes, cast, char, char-array, char?, chars, chunk, %
    chunk-append, chunk-buffer, chunk-cons, chunk-first, chunk-next, chunk-rest, %
    chunked-seq?, class, class?, clear-agent-errors, clojure-version, coll?, %
    commute, comp, comparator, compare, compare-and-set!, compile, complement, %
    concat, conj, conj!, cons, constantly, construct-proxy, contains?, count, %
    counted?, create-ns, create-struct, cycle, dec, dec', decimal?, delay?, %
    deliver, denominator, deref, derive, descendants, destructure, disj, disj!, %
    dissoc, dissoc!, distinct, distinct?, doall, dorun, double, double-array, %
    doubles, drop, drop-last, drop-while, empty, empty?, ensure, %
    enumeration-seq, error-handler, error-mode, eval, even?, every-pred, every?, %
    ex-data, ex-info, extend, extenders, extends?, false?, ffirst, file-seq, %
    filter, filter-ns-publics, filterv, find, find-keyword, find-ns, %
    find-protocol-impl, find-protocol-method, find-var, first, flatten, float, %
    float-array, float?, floats, flush, fn?, fnext, fnil, force, format, %
    frequencies, future-call, future-cancel, future-cancelled?, future-done?, %
    future?, gensym, get, get-in, get-method, get-proxy-class, %
    get-thread-bindings, get-validator, group-by, hash, hash-combine, hash-map, %
    hash-ordered-coll, hash-set, hash-unordered-coll, identical?, identity, %
    ifn?, in-ns, inc, inc', init-proxy, instance?, int, int-array, integer?, %
    interleave, intern, interpose, into, into-array, ints, isa?, iterate, %
    iterator-seq, juxt, keep, keep-indexed, key, keys, keyword, keyword?, last, %
    line-seq, list, list*, list?, load, load-file, load-reader, load-string, %
    loaded-libs, long, long-array, longs, macroexpand, macroexpand-1, %
    make-array, make-hierarchy, map, map-indexed, map?, mapcat, mapv, max, %
    max-key, memoize, merge, merge-with, meta, method-sig, methods, min, %
    min-key, mix-collection-hash, mod, munge, name, namespace, namespace-munge, %
    neg?, newline, next, nfirst, nil?, nnext, not, not-any?, not-empty, %
    not-every?, not=, ns-aliases, ns-functions, ns-imports, ns-interns, %
    ns-macros, ns-map, ns-name, ns-publics, ns-refers, ns-resolve, ns-unalias, %
    ns-unmap, nth, nthnext, nthrest, num, number?, numerator, object-array, %
    odd?, parents, partial, partition, partition-all, partition-by, pcalls, %
    peek, persistent!, pmap, pop, pop!, pop-thread-bindings, pos?, pr, pr-str, %
    prefer-method, prefers, print, print-ctor, print-simple, print-str, printf, %
    println, println-str, prn, prn-str, promise, proxy-call-with-super, %
    proxy-mappings, proxy-name, push-thread-bindings, quot, rand, rand-int, %
    rand-nth, range, ratio?, rational?, rationalize, re-find, re-groups, %
    re-matcher, re-matches, re-pattern, re-seq, read, read-line, read-string, %
    realized?, record?, reduce, reduce-kv, reduced, reduced?, reductions, ref, %
    ref-history-count, ref-max-history, ref-min-history, ref-set, refer, %
    release-pending-sends, rem, remove, remove-all-methods, remove-method, %
    remove-ns, remove-watch, repeat, repeatedly, replace, replicate, require, %
    reset!, reset-meta!, resolve, rest, restart-agent, resultset-seq, reverse, %
    reversible?, rseq, rsubseq, satisfies?, second, select-keys, send, send-off, %
    send-via, seq, seq?, seque, sequence, sequential?, set, %
    set-agent-send-executor!, set-agent-send-off-executor!, set-error-handler!, %
    set-error-mode!, set-validator!, set?, short, short-array, shorts, shuffle, %
    shutdown-agents, slurp, some, some-fn, some?, sort, sort-by, sorted-map, %
    sorted-map-by, sorted-set, sorted-set-by, sorted?, special-symbol?, spit, %
    split-at, split-with, str, string?, struct, struct-map, subs, subseq, %
    subvec, supers, swap!, symbol, symbol?, take, take-last, take-nth, %
    take-while, test, the-ns, thread-bound?, to-array, to-array-2d, trampoline, %
    transient, tree-seq, true?, type, unchecked-add, unchecked-add-int, %
    unchecked-byte, unchecked-char, unchecked-dec, unchecked-dec-int, %
    unchecked-divide-int, unchecked-double, unchecked-float, unchecked-inc, %
    unchecked-inc-int, unchecked-int, unchecked-long, unchecked-multiply, %
    unchecked-multiply-int, unchecked-negate, unchecked-negate-int, %
    unchecked-remainder-int, unchecked-short, unchecked-subtract, %
    unchecked-subtract-int, underive, unsigned-bit-shift-right, update-in, %
    update-proxy, use, val, vals, var-get, var-set, var?, vary-meta, vec, %
    vector, vector-of, vector?, with-bindings*, with-meta, with-redefs-fn, %
    xml-seq, zero?, zipmap}, %
  morekeywords=[5]{def-cps-fn, defanglican, defm, defquery, defun, defproc, defdist}, %
  morekeywords=[6]{cps-fn, fm, lambda, mem, query, with-primitive-procedures}, %
  morekeywords=[7]{%
    doquery, %
    conditional, %
    collect-by, equalize, exec, infer, log-marginal, print-predicts, %
    rand, rand-int, rand-nth, rand-roulette, stripdown, warmup, %
    ->CRP-process, ->DP-process, ->GP-process, %
    ->bernoulli-distribution, ->beta-distribution, ->binomial-distribution, %
    ->categorical-crp-distribution, ->categorical-distribution, %
    ->categorical-dp-distribution, ->chi-squared-distribution, %
    ->dirichlet-distribution, ->discrete-distribution, %
    ->exponential-distribution, ->flip-distribution, ->gamma-distribution, %
    ->mvn-distribution, ->normal-distribution, ->poisson-distribution, %
    ->sample, ->observe, sample*, observe*, %
    ->uniform-continuous-distribution, ->uniform-discrete-distribution, %
    ->wishart-distribution, CRP, DP, GP, abs, absorb, acos, asin, atan, %
    bernoulli, beta, binomial, categorical, categorical-crp, categorical-dp, %
    cbrt, ceil, chi-squared, cos, cosh, cov, dirichlet, discrete, exp, %
    exponential, flip, floor, gamma, gen-matrix, log, log-gamma-fn, %
    log-mv-gamma-fn, log-sum-exp, map->CRP-process, map->DP-process, %
    map->GP-process, map->bernoulli-distribution, map->beta-distribution, %
    map->binomial-distribution, map->categorical-crp-distribution, %
    map->categorical-distribution, map->categorical-dp-distribution, %
    map->chi-squared-distribution, map->dirichlet-distribution, %
    map->discrete-distribution, map->exponential-distribution, %
    map->flip-distribution, map->gamma-distribution, map->mvn-distribution, %
    map->normal-distribution, map->poisson-distribution, %
    map->uniform-continuous-distribution, map->uniform-discrete-distribution, %
    map->wishart-distribution, mvn, normal, poisson, pow, produce, %
    rint, round, signum, sin, sinh, sqrt, tag, tan, tanh, transform-sample, %
    uniform-continuous, uniform-discrete, wishart, %
    add-log-weight, add-predict, clear-predicts, get-log-weight, %
    get-mem, get-predicts, in-mem?, set-log-weight, set-mem, %
  }, %
  morekeywords=[8]{factor, observe, predict, retrieve, sample, store}, %
  sensitive, %
  alsoletter={:,-,+,*,?,/,!,>,<}, %
  morecomment=[l];, %
  morestring=[b]", %
  keywordsprefix=:, %
}[keywords,comments,strings]
 
\cbcolor{red}
% ...

\title{Design and Implementation of Probabilistic Programming Language Anglican}
\authorinfo{David Tolpin \and Jan-Willem van de Meent \and Hongseok Yang \and Frank Wood}{University of Oxford}{dtolpin@robots.ox.ac.uk \and jwvdm@robots.ox.ac.uk \and hongseok.yang@cs.ox.ac.uk \and fwood@robots.ox.ac.uk}

\begin{document}

\maketitle

\begin{abstract}
    Anglican is a probabilistic programming system designed to interoperate
    with Clojure and other JVM languages. We introduce the
    programming language Anglican, outline our design choices,
    and discuss in depth the implementation of the Anglican language
    and runtime, including macro-based compilation, extended
    CPS-based evaluation model, and functional representations for
    probabilistic paradigms, such as a distribution,
    a random process, and an inference algorithm.

    We show that a probabilistic functional language can be
    implemented efficiently and integrated tightly with a
    conventional functional language with only moderate
    computational overhead. We also demonstrate how advanced
    probabilistic modeling concepts are mapped naturally
    to the functional foundation.
\end{abstract}

\section{Introduction}
\label{sec:intro}

For data science practitioners, statistical inference is typically
just one step in a more elaborate analysis workflow. The first stage
of this work involves data acquisition, pre-processing and cleaning.
This is often followed by several iterations of exploratory model
design and testing of inference algorithms. Once a sufficiently
robust statistical model and a corresponding inference algorithm have
been identified, analysis results must be post-processed, visualized,
and in some cases integrated into a wider production system.

Probabilistic programming systems \cite{GMR+08,MSP14,WVM14,GS15}
represent generative models as programs written in a specialized
language that provides syntax for the definition and conditioning of
random variables. The code for such models is generally concise,
modular, and easy to modify or extend. Typically inference can be
performed for any probabilistic program using one or more generic
inference techniques provided by the system backend, such as
Metropolis-Hastings \cite{WSG11,MSP14,YHG14}, Hamiltonian Monte Carlo
\cite{SDT04}, expectation propagation \cite{MWG+10}, and extensions
of Sequential Monte Carlo \cite{WVM14,MYM+15,PWD+14} methods.
These generic techniques may be less statistically efficient
than techniques tailored to a specific model. However,
probabilistic programming facilitates simpler implementations of
models making the inference inherently faster.

While probabilistic programming systems shorten the iteration cycle in
exploratory model design, they typically lack basic functionality needed for
data I/O, pre-processing, and analysis and visualization of inference results.
In this paper, we describe the implementation of
Anglican~\cite{TMW15, Anglican}, a probabilistic
programming language that tightly integrates with
Clojure~\cite{H08,Clojure}, a general-purpose programming language
that runs on the Java Virtual Machine (JVM). Both languages share a common
syntax, and can be invoked from each other. This allows Anglican programs to
make use of a rich set of libraries written in both Clojure and Java.
Conversely, Anglican allows intuitive and compact specification of models for
which inference may be performed as part of a larger Clojure project.

There are several ways to build a programming language
on top of or besides another language.  The easiest is an
interpreter --- a program that reads a program, in its entirety
or line-by-line, and executes it by applying operational
semantics of a certain kind to the language. \textsc{Basic} is
famous for line-by-line interpreted implementations.

Another approach is to write a compiler, either to a virtual
architecture, so called p-code or byte-code, or to real
hardware. Here, the whole program is translated from the
`higher-level' source language to a `lower-level' object
language, which can be directly executed, either by hardware or
by an interpreter --- but the latter interpreter can be made
simpler and more efficient  than an interpreter for the source
language.

On top of these two approaches are methods in which a new
language is implemented `inside' another language of the same
level of abstraction. Different languages provide different
means for this; Lisp is famous for the macro facility
that allows to extend the language almost without
restriction --- by writing \textit{macros}, one adds new
constructs to the existing language. There are several uses of
macros --- one is to extend the language \textit{syntax}, for
example, by adding new control structures; another is to keep
the existing syntax but alter the \textit{operational semantics}
--- the way programs are executed and compute their outputs.

Anglican is implemented in just this way --- a macro facility
provided by Clojure, a Lisp dialect, is used both to extend
Clojure with constructs that delimit probabilistic code, and to
alter the operational semantics of Clojure expressions inside
probabilistic code fragments. Anglican claims its right to count
as a separate language because of the ubiquitous probabilistic
execution semantics rather than a different syntax,
which is actually an advantage rather than a drawback ---
Clojure programmers only need to know how to specify the
boundaries of Anglican programs, but can use familiar Clojure
syntax to write probabilistic code.

An implementation of Anglican must therefore address three issues:
\begin{itemize}
    \item the Clojure syntax to introduce probabilistic Anglican
        code inside Clojure modules;
    \item source-to-source transformation of Anglican programs
        into Clojure, so that probabilistic execution becomes
        possible;
    \item algorithms which run Clojure code, obtained by
        transforming Anglican programs, according to the
        probabilistic operational semantics.
\end{itemize}
Execution of probabilistic programs by inference algorithms is
different from execution of deterministic programs.  A
probabilistic program is executed multiple times, often hundreds
of thousands or even millions of times for a single inference
task. Random choices may affect which parts of the program code
are executed and how often. Many inference algorithms require
re-running the program multiple times partially, from a certain
point on.  Different executions may employ different random
choices. However, for efficient inference a correspondence
between random choices in different executions should be
maintained. These are just some of the challenges which were
faced and solved during development of Anglican.

Comparisons of Anglican with other implementations of
probabilistic programming languages~\cite{SGG15}\cite[pp.
32--33]{P16} demonstrate that Anglican achieves state-of-the-art
computational efficiency without sacrificing expressiveness.
Anglican language syntax, compilation, invocation, and runtime
support of Anglican queries are discussed in detail in further
sections.

\section{Design Outline}
\label{sec:design}

An Anglican program, or \textit{query}, is compiled into a
Clojure function. When inference is performed with a provided
algorithm, this produces a sequence of samples. Anglican shares
a common syntax with Clojure; Clojure functions can be called
from Anglican code and vice versa. A simple program in Anglican
can look like the following code:
\begin{lstlisting}[style=default]
(defquery model data
  "chooses a distribution
   which describes the data"
  (let [;;; Guess a distribution.
        dist (sample (categorical
                       [[normal 0.5]
                        [gamma 0.5]]))
        a (sample (gamma 1 1))
        b (sample (gamma 1 1))
        d (dist a b)]

    ;;; Observe samples from the distribution.
    (loop [observations data]
      (when (not-empty observations)
        ;; Retrieve the first observation as `o',
        ;; and store the rest of observations in 
        ;; 'observations*'.
        (let [[o & observations*] observations]
          ;; Observe 'o' from the guessed 
          ;; distribution 'd'.
          (observe d o))
        ;; Proceed to the next iteration with
        ;; the rest of observations.
        (recur observations*)))

    ;;; Return the distribution and parameters.
    [d a b]))
\end{lstlisting}
The query builds a model for the input $\texttt{data}$, 
a sequence of data points.
It defines a probability distribution on three variables, 
$\texttt{d} \in \{\texttt{normal},\texttt{gamma}\}$
for a distribution type, and $\texttt{a}$ and $\texttt{b}$ for
positive parameters for the type. Concretely,
using the $\texttt{sample}$ forms, the query
first defines a so called prior distribution on these three variables,
and then it adjusts this prior distribution based on observations in
$\texttt{data}$ using the \texttt{observe} form.
Samples from this adapted distribution (also called posterior distribution)
can be obtained by running the query under one of Anglican's
inference algorithms.

Internally, an Anglican query is represented by a computation in
\textit{continuation passing style} (CPS)~\cite{AJ89}, and inference algorithms
exploit the CPS structure of the code to intercept probabilistic
operations in an algorithm-specific way\footnote{\cite{GS15}
also describe a CPS-based implementation of a probabilistic
programming language.}. Among the available
inference algorithms there are Particle Cascade~\cite{PWD+14},
Lightweight Metropolis-Hastings~\cite{WSG11}, Iterative Conditional
Sequential Monte-Carlo (Particle Gibbs)~\cite{WVM14}, and others.
Inference on Anglican queries generates a lazy sequence of
samples, which can be processed asynchronously in Clojure
code for analysis, integration, and decision making.

Clojure (and Anglican) runs on the JVM and gets access to a wide choice of Java
libraries for data processing, networking, presentation, and imaging.
Conversely, Anglican queries can be called from Java and other JVM languages.
Programs involving Anglican queries can be deployed as JVM
\textit{jars}, and run without modification on any platform for
which JVM is available.

A probabilistic program, or query, mostly runs deterministic
code, except for certain checkpoints, in which probabilities are
involved, and normal, linear execution of the program is
disrupted. In Anglican and similar languages there are two types
of such checkpoints:

\begin{itemize}
    \item drawing a value from a random source (\texttt{sample});
    \item conditioning a computed value on a random source
        (\texttt{observe}).
\end{itemize}

Anglican can be mostly implemented as a regular programming
language, except for the handling of these checkpoints.
Depending on the \textit{inference algorithm}, \texttt{sample}
and \texttt{observe} may result in implicit input/output
operations and control changes. For example, \texttt{observe} in
particle filtering inference algorithms \cite{WVM14} is a non-deterministic
control statement at which a particle (corresponding to
a user-level thread executing a program)
can be either replicated or terminated. Similarly, in 
Metropolis-Hastings \cite{WSG11},
\texttt{sample} is both an input
statement which `reads' values from a random source,
and a non-deterministic control statement
(with delayed effect), eventually affecting acceptance or
rejection of a sample.

Because of the checkpoints, Anglican programs must allow the
inference algorithm to step in, recording information and
affecting control flow. This can be implemented through
coroutines/cooperative multitasking, and parallel execution/preemptive
multitasking, as well as through explicit
maintenance of program continuations at checkpoints. Anglican
follows the latter option. Clojure is
a functional language, and continuation-passing style (CPS)
transformation is a well-developed technique in the area of
functional languages. Implementing a variant of CPS
transformation seemed to be the most flexible and lightweight
option --- any other form of concurrency would put a higher burden
on the underlying runtime (JVM) and the operating system.
Consequently, Anglican has been implemented as a CPS-transformed
computation with access to continuations in probabilistic
checkpoints. The Anglican `compiler', represented by a set of
functions in the \texttt{anglican.trap} namespace, accepts a
Clojure subset and transforms it into a variant of CPS
representation, which allows inference algorithms to intervene
in the execution flow at probabilistic checkpoints.

Anglican is intended to co-exist with Clojure and be a part of
the source of a Clojure program. To facilitate this, Anglican
programs, or queries, are wrapped by macros (defined in the
\texttt{anglican.emit} namespace), which call the CPS
transformations and define Clojure values
suitable for passing
as arguments to inference algorithms (\texttt{defquery},
\texttt{query}). In addition to defining entire queries,
Anglican promotes modularization of inference algorithms through
the definitions of \textit{probabilistic functions} using
\texttt{defm}
(Anglican counterparts of Clojure \texttt{defn}).
Probabilistic functions are
written in Anglican, may include probabilistic forms
\texttt{sample} and \texttt{observe}, and can be seamlessly
called from inside Anglican queries, just like functions locally
defined within the same query.

Operational semantics of Anglican queries is different from that
of Clojure code, therefore queries must be called through
inference algorithms, rather than `directly'.  The
\texttt{anglican.inference} namespace declares the (ad-hoc) polymorphic
function \texttt{infer} using Clojure's multimethod mechanism.
This function accepts an Anglican query and
returns a lazy sequence of weighted samples from the
distribution defined by the query.  When inference
is performed on an Anglican query, the query is run by a
particular inference algorithm. Inference algorithms must
provide an implementation for \texttt{infer}, as well as
override the polymorphic function \texttt{checkpoint} (defined
as a multimethod) so as to handle \texttt{sample} and
\texttt{observe} in an algorithm-specific manner and
to construct an appropriate result on the
termination of a probabilistic program.

Finally, Anglican queries use `primitive', or commonly known
and used, distributions, to draw random samples and condition
observations. Many primitive distributions are provided by the
\texttt{anglican.runtime} namespace, and an additional
distribution can be defined by the user by implementing a
particular set of functions for the distribution
(via Clojure's protocol mechanism). The \texttt{defdist} macro
provides a convenient syntax for defining primitive distributions.

\section{Language}
\label{sec:language}

The Anglican language is a subset of Clojure\footnote{It would be
possible to support almost full Clojure by expanding all macros
in the Anglican source code. However, in Clojure, unlike in
Scheme~\cite{SDF+10} or Common Lisp~\cite{PC94}, the result
of macro-expansion of derived special forms is not well
specified and implementation specific.}.  Anglican queries
are defined within \texttt{defquery} as shown in the previous section,
using \texttt{if}, \texttt{when}, \texttt{cond}, \texttt{case},
\texttt{let}, \texttt{and}, \texttt{or}, \texttt{fn} forms;
other forms of Clojure may be supported in the future but are not now.
In \texttt{let} bindings and \texttt{fn} argument lists,
Clojure's pattern matching mechanism for vector data type
(called vector destructuring) is supported. Also, compound
literals for vectors, hash maps, and
sets are supported just like in Clojure.

\subsection{Core Library}
\label{sec:core}

All of Clojure's core library except for higher-order functions
(functions that accept other functions as arguments) is
available in Anglican. Higher-order functions cannot be reused
from Clojure, as they have to be re-implemented to accept
functional arguments in CPS form. The following higher-order
functions are implemented: \texttt{map}, \texttt{reduce},
\texttt{filter}, \texttt{some}, \texttt{repeatedly},
\texttt{comp}, \texttt{partial}.

\subsection{Tail Call}
\label{sec:tail}

Clojure provides special forms \texttt{loop} and \texttt{recur}
for writing tail-recursive programs. Such forms are not necessary
in Anglican, because Anglican programs are CPS-converted and do not
use the stack. For instance, no recursive call in Anglican can lead to stack overflow.
In fact, in Anglican, it is recommended to use recursive calls to functions instead of
\texttt{recur}. However, \texttt{loop}/\texttt{recur} is provided in Anglican
for convenience as a way to express loops. \texttt{recur} outside of \texttt{loop}
will lead to unpredictable behaviour and hard-to-catch errors.

\section{Macro-based Compilation}
\label{sec:compilation}

Compilation of Anglican into Clojure is built around a variant of CPS
transformation. In a basic CPS-transformed program, a
continuation receives a single argument --- the computed value.
In Anglican, there are two flows of computation going in
parallel: values are computed by functional code, and,
at the same time, the state of the probabilistic program,
used by inference algorithms, is updated by
probabilistic forms. Because of that, in Anglican a continuation
accepts  \textit{two} arguments:
\begin{itemize}
    \item the computed value;
    \item the internal state, bound to the local variable
        \texttt{\$state} in every lexical scope.
\end{itemize}

The compilation relies on the Clojure \textit{macro} facility,
and implemented as a library of \textit{functions} in namespace
\texttt{anglican.trap}, which are invoked by macros.  The CPS
transformation is organized in top-down manner.  The top-level
function is  \texttt{cps-of-expression}, which receives an
expression and a continuation, and returns the expression in the
CPS form.  For example, the CPS transformation of constant
\texttt{1} with continuation \texttt{cont} thus takes the
following form:
\begin{lstlisting}[style=default]
=> (cps-of-expression 1 'cont)
(cont 1 $state)
\end{lstlisting}

\subsection {The State}
\label{sec:state}

The state (\texttt{\$state}) is threaded through the computation
and contains data used by inference algorithms. \texttt{\$state}
is a Clojure hash map:
\begin{lstlisting}[style=default]
(def initial-state
  "initial program state"
  {:log-weight 0.0,
   :result nil,
   ::mem {},
   ::store nil,
   ... })
\end{lstlisting}
which records inference-relevant information under various keys
such as $\texttt{:log-weight}$ and $\texttt{::mem}$.
The full list of map entries depends on the inference
algorithm. Except for transformation of
the \texttt{mem} form (which converts a function to one with memoization), 
CPS transformation routines
are not aware of contents of \texttt{\$state}, do not access or
modify it directly, but rather just thread the state unmodified
through the computation. Algorithm-specific handlers of
checkpoints corresponding to the probabilistic forms
(\texttt{sample}, \texttt{observe}) modify the
state and reinject a new state into the computation.

\subsection{Expression Kinds}
\label{sec:expr}

There are three different kinds of inputs to CPS transformation:
\begin{itemize}
    \item Literals, which are constant expressions. They
       are passed as an argument to the
        continuation unmodified.
    \item Value expressions such as the \texttt{fn} form
        (called \textit{opaque}
        expressions in the code). They must be transformed to
        CPS, but the transformed object is passed to the
        continuation as a whole, opaquely.
    \item General expressions (which we call
        \textit{transparent} expressions).
        The continuation is threaded through such an expression
        in an expression-specific way, and can be
        called in multiple locations of the CPS-transformed
        code, such as in all branches of an \texttt{if}
        statement.
\end{itemize}

\subsubsection{Literals}
\label{sec:literals}

Literals are the same in Anglican and Clojure. They are
left unmodified; literals are a subset of opaque expressions.
However, the Clojure syntax has a peculiarity
of using the syntax of compound literals (vectors, hash maps,
and sets) for data constructors. Hence, compound literals must
be traversed recursively, and if there is a nested non-literal
component, transformed into a call to the corresponding data
constructor. Functions \texttt{cps-of-vector},
\texttt{cps-of-hash-map}, \texttt{cps-of-set}, called from
\texttt{cps-of-expression}, transform Clojure constructor syntax
(\texttt{[...]}, \texttt{\{...\}}, \texttt{\#\{...\}}) into the
corresponding calls:
\begin{lstlisting}[style=default]
=> (cps-of-vector [0 1 2] 'cont)
(cont (vector 0 1 2) $state)
=> (cps-of-hash-map {:a 1, :b 2} 'cont)
(cont (hash-map :a 1, :b 2) $state)
=> (cps-of-set #{0 1} 'cont)
(cont (set (list 0 1)) $state)
\end{lstlisting}

\subsubsection{Opaque Expressions}
\label{sec:opaque}

Opaque, or value, expressions, have a different shape in the
original and the CPS form. However, their CPS form follows the
pattern \texttt{(continuation transformed-expression \$state)}, and thus
the transformation does not depend on the continuation parameter, and
can be accomplished without passing the parameter as a
transformation argument. Primitive (non-CPS) procedures used in
Anglican code, \texttt{(fn ...)} forms, and \text{(mem ...)}
forms are opaque and transformed by
\texttt{primitive-}\linebreak[0]\texttt{procedure-}\linebreak[0]\texttt{cps},
\texttt{fn-cps}, and \texttt{mem-cps}, correspondingly: a
slightly simplified CPS form of expression
\begin{lstlisting}[style=default]
(fn [x y]
  (+ x y))
\end{lstlisting}
would be
\begin{lstlisting}[style=default]
(fn [cont $state x y]
  (cont (+ x y) $state))
\end{lstlisting}
In the actual code an automatically generated fresh symbol is
used instead of \texttt{cont}.

\subsubsection{General Expressions}
\label{seq:general}

The most general form of CPS transformation receives an
expression and a continuation as parameters, and returns the expression
in CPS form with the continuation parameter potentially called in multiple
tail positions. General expressions can be somewhat voluntarily
divided into several groups:
\begin{itemize}
    \item binding forms --- \texttt{let} and
        \texttt{loop/recur};
    \item flow control --- \texttt{if}, \texttt{when},
        \texttt{cond}, \texttt{case}, \texttt{and}, \texttt{or} and \texttt{do};
    \item function applications and \texttt{apply};
    \item probabilistic forms --- \texttt{observe},
        \texttt{sample}, \texttt{store}, and \texttt{retrieve}.
\end{itemize}
Functions that transform general expressions accept the
expression and the continuation as parameters, and are
consistently named \texttt{cps-of-}\textit{form}, for example,
\texttt{cps-of-do}, \texttt{cps-of-store}.

\subsection{Implementation Highlights}
\label{seq:highlights}

So far we introduced the basics of Anglican compilation to
Clojure. The described approaches and techniques are important
for grasping the language implementation but relatively well-known.
In the rest of the section we focus on challenges we met and
resolved while implementing Anglican, as well as on
implementation of unique features of Anglican as a probabilistic
programming language.

\subsubsection{Continuations}
\label{seq:continuations}

Continuations are functions that are called in tail positions
with the computed value and state as their arguments --- in CPS
there is always a function call in every tail position and never
a value. Continuations are passed to CPS transformers, and when
transformers are called recursively, the continuations are
generated on the fly.

There are two critical issues related to generation of
continuations:
\begin{itemize}
    \item unbounded \textit{stack growth} in recursive code;
    \item code size \textit{explosion} when a non-atomic
        continuation is symbolically substituted in multiple
        locations.
\end{itemize}

\paragraph{Managing stack size}

In implementations of functional programming languages stack
growth is avoided through \textit{tail call optimization} (TCO).
However, Clojure does not support a general form of TCO, and
CPS-transformed code that creates deeply nested calls will
easily exhaust the stack. Anglican employs a workaround called
\textit{trampolining} --- instead of inserting a continuation
call directly, the transformer always wraps the call into a
\textit{thunk}, or parameterless function. The thunk is returned
and called by the trampoline (Clojure provides function
\texttt{trampoline} for this purpose) --- this way the
computation continues, but the stack is collapsed on every
continuation call. Function \texttt{continue} implements the
wrapping and is invoked on every continuation call:
\begin{lstlisting}[style=default]
=> (continue 'cont 'value 'state)
(fn [] (cont value state))
\end{lstlisting}
Correspondingly, the full, wrapped CPS form of
\begin{lstlisting}[style=default]
(fn [x y] (+ x y))
\end{lstlisting}
becomes
\begin{lstlisting}[style=default]
(fn [cont $state x y]
  (fn []
    (cont (+ x y) $state)))
\end{lstlisting}
When the CPS-transformed function is called, it returns a
\textit{thunk} (a parameterless function) which is then
re-invoked through the trampoline, with the stack collapsed.

\paragraph{Avoiding exponential code growth}

To realize potential danger of code size explosion, consider
CPS transformation of code
\begin{lstlisting}[style=default]
(if (adult? person)
  (if (male? person)
     (choose-beer)
     (choose-wine))
  (choose-juice))
\end{lstlisting}
with continuation
\begin{lstlisting}[style=default]
(fn [choice _]
  (case (kind choice)
    :beer (beer-jar choice)
    :wine (wine-glass choice)
    :juice (juice-bottle choice)))
\end{lstlisting}
During CPS transformation, if we substitute the code of this continuation 
for all of its calls, the code will be repeated three times in 
the CPS-transformed expression. In general, CPS code can
grow extremely large if the code of continuations is substituted
repeatedly.

To circumvent this inefficiency, CPS transformers for
expressions with multiple continuation points (\texttt{if} and
derivatives, \texttt{and}, \texttt{or}, and \texttt{case}) bind
the continuation to a fresh symbol if it is not yet a symbol.
Macro \texttt{defn-with-{\linebreak[0]}named-{\linebreak[0]}cont}
establishes the binding automatically:
\begin{lstlisting}[style=default]
=> (cps-of-if '(c t f) '(fn [x] (* 2 x)))
(let [cont (fn [x] (* 2 x))]
  (if c
      (fn [] (cont t $state))
      (fn [] (cont f $state))))
\end{lstlisting}

\subsubsection{Primitive Procedures}
\label{seq:primitive}

When an Anglican function is transformed into a Clojure function
by \texttt{fn-cps}, two auxiliary parameters are added to the
beginning of the parameter list --- continuation and state.
Correspondingly, when a function \textit{call} is transformed
(by \texttt{cps-of-application} or \texttt{cps-of-apply}), the
current continuation and the state are passed to the called
function. Anglican can also call Clojure functions; however,
Clojure functions do not expect these auxiliary parameters.
To allow the mixing of Anglican (CPS-transformed) and Clojure
function calls in Anglican code, the Anglican compiler must be
able to recognize `primitive' (that is, implemented in Clojure
rather than in Anglican) functions.

Providing an explicit syntax for differentiating between
Anglican and Clojure function calls would be cumbersome. Another
option would be to use meta-data to identify Anglican function
calls at runtime. However, this would impact performance, and
good runtime performance is critical for probabilistic
programs. The approach taken by Anglican is to maintain a list
of unqualified names of primitive functions, as well of
namespaces in which all functions are primitive, and recognize
primitive functions by name --- if a function name is not in the
list, the function is an Anglican function. Global dynamically-bound 
variables \texttt{*primitive-procedures*} and
\texttt{*primitive-namespaces*} contain the initial lists of
names and namespaces, correspondingly. Of course, local bindings
can shade global primitive function names. For example,
\texttt{first} is a Clojure primitive function that takes the first
element from an ordered collection such as list and vector,
but inside the let block in the
following example, \texttt{first} is an Anglican function:
\begin{lstlisting}[style=default]
(let [first (fn [[x & y]] x)]
  (first '[1 2 3]))
\end{lstlisting}
The Anglican compiler takes care of the shading by rebinding
\texttt{*primitive-procedures*} in every lexical scope
(\texttt{fn-cps}, \texttt{cps-of-let}). Macro
\texttt{shading-primitive-procedures} automates the shading.

\subsubsection{Probabilistic Forms}
\label{seq:forms}

There are two proper probabilistic forms turning Anglican into a
probabilistic programming language --- \texttt{sample} and
\texttt{observe}. Their purpose is to interrupt deterministic
computation and transfer control to the inference algorithm.
Practically, this is achieved through returning
\textbf{checkpoints} --- Clojure values of the corresponding
types (\texttt{anglican.trap.sample} or
\texttt{anglican.trap.observe}). The types contain fields
specific to each form, as well as the continuation; calling the
continuation resumes the computation. Checkpoints expose the
program state to the inference algorithm, and the updated state
is re-injected into the computation when the continuation is
called:
\begin{lstlisting}[style=default]
=> (cps-of-expression '(sample dist) 'cont)
(->sample dist cont $state)
=> (cps-of-expression '(observe dist v) 'cont)
(->observe dist v cont $state)
\end{lstlisting}
Here \texttt{->sample} and \texttt{->observe} in the CPS-transformed
expressions are constructors for Clojure records, and they take 
values of their fields as arguments.

In addition to checkpoints, there are a few other special forms
---  \texttt{store}, \texttt{retrieve}, \texttt{mem} --- which
modify program state. These forms are translated into
expressions involving calls of functions from the
\texttt{anglican.state} namespace. The \texttt{mem} form, which
implements memoization, deserves a more detailed explanation.

\paragraph{Memoization}
The author of a probabilistic model might want to 
randomly draw a feature value for each entity in a collection,
but to retain the same drawn
value for the same entity during a single run of the probabilistic
program. For example, a person may have brown or green eyes
with some probability, but the \textit{same} person will
always have the same eye color. This can be achieved
through the use of memoized functions. In Anglican, one
might write:
\begin{lstlisting}[style=default]
(let [eye-color (mem (fn [person]
                       (sample
                         (categorical
                           ['brown 0.5]
                           ['green 0.5]))))]
  (if (not= (eye-color 'bill) (eye-color 'john))
    (eye-color 'bill)
    (eye-color 'john)))
\end{lstlisting}
The \texttt{mem} form converts a function to a memoized variant,
which remembers past inputs and the corresponding outputs, and 
returns the remembered output when given such a past input, 
instead of calling the original function with it. As a result,
for every input, random draws will be made only for the first time 
that the memoized function is called with the input; all subsequent
calls with the input will just reuse these draws and return the same output.

Memoization is often implemented on top of a mutable
dictionary, where the key is the argument list and the value is
the returned value. However, there are no mutable data
structures in a probabilistic program. Hence, \texttt{mem}'s
memory is stored as a nested dictionary in the program state
introduced during CPS transformation (function \texttt{mem-cps}).  
Every memoized function gets a
unique automatically generated identifier. Each time a memoized
function is called, one of two continuations is chosen,
depending on whether the same function (a function with the same
identifier) was previously called in the same run of the
probabilistic program with the same arguments. If the memoized
result is available, the continuation of the memoized function
call is immediately called with the stored result. Otherwise,
the argument of \texttt{mem} is called with a continuation which
first creates an updated state with the memoized result, and
then calls the `outer' continuation with the result and the
updated state:
\begin{lstlisting}[style=default]
=> (mem-cps '(foo))
(let [M (gensym "M")]
  (fn mem23623 [C $state & P]
    (if (in-mem? $state M P)
      ;; previously memoized result
      (fn []
        (C (get-mem $state M P) $state))
      ;; new computation
      (clojure.core/apply
        foo
        ;; memoize result in state
        (fn [V $state]
          (fn [] (C V (set-mem $state M P V))))
        $state
        P)))))
\end{lstlisting}

Memoized results are not shared among multiple runs of a
probabilistic program, which is intended. Otherwise, it would be
impossible to memoize functions with random results.

\section{Inference Algorithms}
\label{sec:inference}

A probabilistic program in Anglican may look almost like a
Clojure program. However, the purpose of executing a
probabilistic program is different from that of a `regular'
program: instead of producing the result of a single execution,
a probabilistic program computes, exactly or approximately, the
distribution from which execution results are drawn.  Computing
the distribution is facilitated by an inference algorithm.

Probabilistic programming system Anglican provides a variety of
approximate inference algorithms. Ideally, Anglican should
automatically choose the most appropriate algorithm for each
probabilistic program. In practice, selecting an inference
algorithm, or a combination thereof, is still a challenging
task, and program structure, intended use of the computation
results, performance, approximation accuracy, and other factors
must be taken into consideration. New algorithms are being
developed and added to Anglican~\cite{TMP+15,MPT+16,RNL+2016},
as a part of ongoing research.

In the implementation of Anglican, inference algorithms are
instantiations of the (ad-hoc) polymorphic function
\texttt{infer} declared as Clojure's multimethod in the
\texttt{anglican.{\linebreak[0]}inference} namespace. The function accepts an
algorithm identifier, a query --- the probabilistic program in
which to perform the inference, an initial value for the query,
and optional algorithm parameters.

\subsection{The \texttt{infer} Function}
\label{sec:infer}

The sole purpose of the algorithm identifier of \texttt{infer} is to invoke
an appropriate overloading or implementation of the function --- conventionally,
the identifier is a Clojure keyword (a symbolic constant starting with colon)
related to the algorithm name, such as \texttt{:lmh} for Lightweight
Metropolis-Hastings and \texttt{:pcascade} for Particle Cascade.
The second parameter is a query as defined by
the \texttt{defquery} form or its anonymous version \texttt{query}. For instance,
the following Clojure code invokes \texttt{infer} on an Anglican query
defined anonymously via the \texttt{query} form:
\begin{lstlisting}[style=default]
(let [x 1]
  (infer :pgibbs (query x) nil))
\end{lstlisting}
A query is executed by calling the initial continuation of the
query, which accepts a value and a state. The state is supplied
by the inference algorithm, while the value is provided as a
parameter of \texttt{infer}. A query does not have to have any
parameters, in which case the value can be simply \texttt{nil}.
When a query is defined with a binding for the initial value,
the value becomes available inside the query. A query may
accept multiple parameters using Clojure's structured binding.
For instance, it may take multiple parameters as components
of an input vector. In this case, the initial value is given
as a structured value, such as a vector, and the components
of this value are matched to the corresponding
parameters of the query via the destructuring mechanism of Clojure.
For example,
\begin{lstlisting}[style=default]
(defquery my-query [mean sd]
  (sample (normal mean sd)))

(def samples (infer :lmh my-query [1.0 3.0]))
\end{lstlisting}
Finally, any number of auxiliary arguments can be
passed to \texttt{infer}.
By convention, the arguments should be
keyword arguments, and are interpreted in the algorithm-specific
manner.

\subsection{Internals of an Inference Algorithm}
\label{seq:internals}

Implementing an inference algorithm in Anglican amounts
to defining an appropriate version of the \texttt{infer} function and 
checkpoint handlers for \texttt{sample} and \texttt{observe}.
The definitions may just reuse default implementations or override them
with new algorithm-specific treatment of \texttt{sample} and \texttt{observe} forms
and inference state.

Let us illustrate this implementation step with
\textit{importance sampling},
the simplest inference algorithm.
Here is an implementation of the \texttt{infer} function for the algorithm:
\begin{lstlisting}[style=default]
(derive ::algorithm
        :anglican.inference/algorithm)
;; invoked when algo parameter is :importance
(defmethod infer :importance 
  [algo prog value & {}]
  (letfn
    ;; recursive function without any parameter
    [sample-seq 
      []
      ;; lazy infinite sequence 
      (lazy-seq
        (cons
          (:state (exec ::algorithm prog value
                        initial-state))
          (sample-seq))))]
    (sample-seq)))
\end{lstlisting}
This implementation is dispatched 
when \texttt{infer} receives \texttt{:importance} as its \texttt{algo} 
parameter. (Anglican includes multiple implementations of \texttt{infer}
for different inference algorithms.) Once dispatched, it lazily 
constructs an infinite sequence of inference states by repeatedly 
executing the program \texttt{prog} using \texttt{exec}
and retrieving the final inference state of the execution 
using \texttt{:state}. For checkpoint handlers, importance sampling simply
relies on their default implementations.

Typically, an inference
algorithm has its own implementations of
\texttt{checkpoint} for \texttt{sample}, \texttt{observe}, or
both, as well as invokes \texttt{exec} from an elaborated
conditional control flow. LMH (\texttt{anglican.lmh}) and SMC
(\texttt{anglican.smc}) are examples of inference algorithms
where either \texttt{observe} (SMC) or \texttt{sample} (LMH)
handler is overridden. In addition, SMC runs multiple particles
(corresponding to user-level threads) simultaneously, while LMH re-runs
programs from an intermediate continuation rather than from the
beginning.

\subsection{Addressing of Checkpoints}
\label{sec:addressing}

Many inference algorithms memoize and reuse earlier computations
at checkpoints. Variants of Metropolis-Hastings reuse previously
drawn values~\cite{WSG11}, as well as additional information
used for adaptive sampling~\cite{TMP+15} at \texttt{sample}
checkpoints. Asynchronous SMC (Particle Cascade) computes
average particle weights at \texttt{observe}
checkpoints~\cite{PWD+14}. Implementations of black-box
variational inference~\cite{WW13,MPT+16} associate with random
variables the learned parameters of variational posterior
distribution.

Checkpoints can be uniquely named at compilation time;
however, at runtime a checkpoint corresponding to a single code
location may occur multiple time due to recurrent invocation of
the function containing the checkpoint. Every unique occurrence
of a checkpoint must receive a different address. An addressing
scheme based on computing of stack addresses of checkpoints was
described in the context of Lightweight
Metropolis-Hastings~\cite{WSG11}.  This scheme has advantages
over the naive scheme where dynamic occurrences of checkpoints
are numbered sequentially. However, it impacts the
performance of probabilistic programs because of the computation
cost of computing the stack addresses:
\begin{enumerate}
    \item On every function call, a component is added to the
        address. Hence, the address size is linear in stack depth.
    \item Every function call must be augmented by symbolic
        information required to compute stack addresses.
\end{enumerate}
In addition, the above scheme can still lead to inferior
correspondence between checkpoints in different traces: in
Anglican and other probabilistic languages where distributions
are first-class objects checkpoints with incompatible arguments
can correspond to the same stack address. Consider the following
program fragment:
\begin{lstlisting}[style=default]
(let [dist (if use-gamma
               (gamma 2. 2.)
               (normal 0. 1.))]
  (sample dist))
\end{lstlisting}
The \texttt{sample} checkpoint has the same stack address in
different traces. However, the random values should not be
reused between different distributions. Further on, in some
algorithms, such as black-box variational inference, the role of
checkpoint addresses is semantic rather than heuristic ---
appropriate correspondence must be established
between checkpoints in different traces for the algorithm to
work.

To overcome the above problems, Anglican uses a different scheme
which is almost as efficient as the scheme based on stack addresses for
reuse of previously drawn values, while producing addresses of constant size,
and allows manual computation of checkpoint addresses at runtime
when the default automatic scheme is insufficient. According to
the scheme:
\begin{itemize}
    \item A checkpoint may accept an auxiliary argument --- the
        checkpoint identifier. If specified, the identifier is the first
        argument of a checkpoint. For example \texttt{(sample
        'x1  (normal 0 1))} defines a \texttt{sample} checkpoint
        with identifier \texttt{x1}.
    \item If the identifier is omitted, a unique identifier is
        generated automatically as a fresh symbol.
    \item At runtime, the address of a checkpoint invocation has
        the form
        \texttt{[}\textit{checkpoint-identifier
        number-of-previous-occurrences}\texttt{]}, where the
        occurrences are of a checkpoint with the same checkpoint
        identifier.
    \item If a sequence of invocations of the same checkpoint is
        interrupted by a different checkpoint, the number of
        previous occurrences is rounded up to a multiple of
        an integer. For efficiency, a small power of 2 is used,
        such as $2^4 = 16$.
\end{itemize}

Consider the following simplified Anglican query:

\begin{lstlisting}[style=default]
(defm foo []
  (if (sample 'C1 (flip 0.5))
    (foo)
    (bar)))

(defm bar []
  (case (sample 'C2 (uniform-discrete 0 3))
    0 (bar)
    1 (foo)
    2 (sample 'C3 (normal 0. 1.))))

(defquery baz
  (foo))
\end{lstlisting}

An execution of the query may result in the following
sequence of checkpoint invocations:
\begin{lstlisting}[style=default]
    (sample 'C1 ...)
    (sample 'C2 ...)
    (sample 'C2 ...)
    (sample 'C1 ...)
    (sample 'C1 ...)
    (sample 'C1 ...)
    (sample 'C2 ...)
    (sample 'C3 ...)
\end{lstlisting}
According to the addressing scheme, the addresses generated
for this invocations are
\begin{lstlisting}[style=default]
    [C1 0]
    [C2 0]
    [C2 1]
    [C1 16]
    [C1 17]
    [C1 18]
    [C2 16]
    [C3 0]
\end{lstlisting}
If the program is run by a Metropolis-Hastings algorithm, then a small change usually takes place in the sequence of checkpoints with each invocation, and the new sequence may become
\begin{lstlisting}[style=default]
    (sample 'C1 ...)
    (sample 'C2 ...)
    (sample 'C1 ...)
    (sample 'C1 ...)
    (sample 'C2 ...)
    (sample 'C2 ...)
    (sample 'C3 ...)
\end{lstlisting}
The addresses for the new sequence are
\begin{lstlisting}[style=default]
    [C1 0]
    [C2 0]
    [C1 16]
    [C1 17]
    [C2 16]
    [C2 17]
    [C3 0]
\end{lstlisting}
Despite the change, the correspondence between checkpoints of
similar types occurring in similar positions (relative positions
in contiguous subsequences of a certain type) is preserved, and
drawn values can be reused efficiently:

\vspace{\baselineskip}
{
\begin{tabular}{l l}
    \textbf{old} & \textbf{new} \\ \hline
    {}\texttt{[C1 0]} & \texttt{[C1 0]} \\
    {}\texttt{[C2 0]} & \texttt{[C2 0]} \\
    {}\texttt{[C2 1]} & \textit{unused} \\
    {}\texttt{[C1 16]} & \texttt{[C1 16]} \\
    {}\texttt{[C1 17]} & \texttt{[C1 17]} \\
    {}\texttt{[C1 18]} & \textit{unused} \\
    {}\texttt{[C2 16]} & \texttt{[C2 16]} \\
    \textit{missing} & \texttt{[C2 17]} \textit{drawn}\\
    {}\texttt{[C3 0]} & \texttt{[C3 0]}
\end{tabular}
\vspace{\baselineskip}}

Note that correspondence between checkpoints in different
traces plays the role of an heuristic in Metropolis-Hastings
family of algorithms. Any correspondence (or no correspondence,
meaning all values must be re-drawn from their proposal
distributions) is valid, and reused values from the previous
invocation which are not in support or have zero probability in
the new invocation are simply ignored and new values are
re-drawn.

This way, each occurrence of a checkpoint has unique address,
but small disturbances --- removal or addition of a single or just
a few checkpoints --- are unlikely to derail the entire sequence.
The probability of derailment depends on the padding. The
padding can be safely, and without any impact on performance,
set to rather large numbers. However, rounding up to a multiple
of 16 proved to be appropriate for all practical purposes.

Function \texttt{checkpoint-id} in the
\texttt{anglican.inference} namespace automates generation of
checkpoint addresses and can be used from any inference
algorithm.

\section{Definitions and Runtime Library}
\label{seq:runtime}

A Clojure namespace that includes a definition of an Anglican
program imports (`requires') two essential namespaces:
\texttt{anglican.emit} and \texttt{anglican.runtime}. The former
provides macros for defining Anglican programs
(\texttt{defquery}, \texttt{query}) and functions
(\texttt{defm}, \texttt{fm}, \texttt{mem}), as well as Anglican
bootstrap definitions that must be included with every program
--- first of all, CPS implementations of higher-order functions.
\texttt{anglican.emit} can be viewed as the Anglican
\textit{compiler tool}, which helps transform Anglican code into
Clojure before any inference is performed.

\texttt{anglican.runtime} is the Anglican runtime library. For
convenience, it exposes common mathematical functions
(\texttt{abs}, \texttt{floor}, \texttt{sin}, \texttt{log},
\texttt{exp}, etc.), but most importantly, it provides
definitions of common distributions. Each distribution object
implements a distribution interface with the \texttt{sample*} and \texttt{observe*}
methods; this interface is defined using Clojure's protocol mechanism
(\texttt{anglican.{\linebreak[0]}runtime/{\linebreak[0]}distribution}).
The \texttt{sample*} \textit{method} returns a random sample and
roughly corresponds to the \texttt{sample} checkpoint,
the \texttt{observe*} \textit{method} returns the log probability
of the value and roughly corresponds to the \texttt{observe}
checkpoint. The methods can be, and sometimes are called from
handlers of the corresponding checkpoints, but do not have to
be. For example, in LMH either the \texttt{sample*} or
the \texttt{observe*} \textit{method} is called for a
\texttt{sample} checkpoint, depending on whether the value is
drawn or reused.

Macro \texttt{defdist} should be used to define distributions.
\texttt{defdist} takes care of defining a separate type
for every distribution so that Clojure multimethods (or overloaded
methods) can be dispatched on
distribution types when needed, e.g. for custom proposal distributions 
used in an inference algorithm. The
Bernoulli distribution could be defined as follows:
\begin{lstlisting}[style=default]
(defdist bernoulli
  "Bernoulli distribution"
  [p]
  (sample* [this] (if (< (rand) p) 1 0))
  (observe* [this value]
    (Math/log (case value
                1 p
                0 (- 1. p)
                0.))))
\end{lstlisting}

In addition to distributions, Anglican provides \textit{random processes},
which define sequences of random variables that are not independent and
identically distributed. Random processes are similar to the so called
`exchangeable random procedures' in Church \cite{GMR+08} and
Venture \cite{MSP14}. However, random sequences generated by Anglican random 
processes are not required to be exchangeable. Random processes are
defined using the \texttt{defproc} macro and implement the
\texttt{anglican.runtime/{\linebreak[0]}random-process} protocol. This
protocol has two methods
\begin{itemize}
\item \texttt{produce}, which returns the distribution on the next
random variable in the sequence, and
\item \texttt{absorb}, which incorporates the value for the next random
variable and returns an updated random process.
\end{itemize}
 As an example, here is a definition of a beta-Bernoulli process, in which
each random variable is distributed according to a Bernoulli distribution with
an unknown parameter that is drawn from a beta distribution:
\begin{lstlisting}[style=default]
(defproc beta-bernoulli
  "beta-Bernoulli process"
  [a b]
  (produce [this] (bernoulli (/ a (+ a b))))
  (absorb [this x]
    (case x
      0 (beta-bernoulli a (inc b))
      1 (beta-bernoulli (inc a) b))))
\end{lstlisting}
Unlike typical implementations of exchangeable random processes, Anglican's random processes do not have mutable state. The \texttt{produce} and \texttt{absorb} methods are deterministic and functional, and therefore do not have corresponding special forms in Anglican. A sequence of random values can be generated using a recursive loop in which \texttt{absorb} returns the updated process for the next iteration. For example:
\begin{lstlisting}[style=default]
(defm sample-beta-binomial [n a b]
  (loop [process (beta-bernoulli a b)
         values []]
    (if (= (count values) n)
      values
      (let [dist (produce process)
            value (sample dist)]
      (recur (absorb process value)
             (conj values value))))))
\end{lstlisting}
Similarly, random processes can also be used to recursively observe a sequence of values:
\begin{lstlisting}[style=default]
(loop [process (beta-bernoulli 1.0 1.0)
       [value0 & values0] values] 
  (let [dist (produce process)]
    (observe dist value0)
    (recur (absorb process value0) values0)))
\end{lstlisting}



\section{Case Study}
\label{seq:study}

The description of Anglican language and environment would be
incomplete without a case study showing the process of building
and executing a solution of an inference problem. This case
study takes a problem for which a solution is not immediately
obvious, \textit{the Deli dilemma}, and guides through
writing an Anglican program for the problem, executing the
program, and post-processing results. 

The program presented in this section is intentionally short and
simple. Anglican is capable of compiling and running elaborated
programs handling large amounts of data. Advanced examples of
Anglican programs and inference can be found in literature on
applications of probabilistic programming~\cite{PLW15, P16, MPT+16}.

\subsection{The Deli Dilemma}
\label{seq:deli}

Imagine that we are facing the following dilemma:
\begin{quote}
A customer wearing round sunglasses came to a deli at 1:13pm,
and grabbed a sandwich and a coffee. Later on the same day, a
customer wearing round sunglasses came at 6:09pm and ordered a
dinner.  Was it the same customer?
\end{quote}
In addition to the story above, we know that:
\begin{itemize}
    \item There is an adjacent office quarter, and it takes
        between 5 and 15 minutes on average to walk from an
        office to the deli, where different average times are  for
        different buildings in the quarter.
    \item Depending on traffic lights, the walking time varies by about 2
        minutes.
    \item The lunch break starts at 1:00pm, and the workday ends at 6:00pm.
    \item The waiter's odds that this is the same customer are 2 to 1.
\end{itemize}

\subsection{Anglican Query}
\label{seq:deli-query}

We want to formalize the dilemma in an Anglican query, based on
the knowledge we have. Let us formalize the knowledge in
Clojure (the times are in minutes). First, we encode our prior
information which holds true independently of the customer's
visit:

\begin{lstlisting}[style=default]
(def mean-time-to-arrive
     "average time to arrive"
     10.)
(def sd-time-to-arrive
     "standard deviation of arrival time"
     3.)
(def time-sd
     "walking time deviation"
     1.)
\end{lstlisting}

Then, we record our observations, based on which we want to
solve the dilemma:

\begin{lstlisting}[style=default]
(def lunch-delay
     "time between lunch break and lunch order"
     13.)
(def dinner-delay
     "time between end of day and dinner order"
     9.)
(def p-same
     "prior probability of the same customer"
     (/ 2. 3.))
\end{lstlisting}

For inference, one often chooses a known distribution to
represent uncertainty. We choose the normal
distribution for representing uncertainty about average arrival
time.

\begin{lstlisting}[style=default]
(def time-to-arrive-prior 
     "prior distribution of average arrival time"
     (normal mean-time-to-arrive
             sd-time-to-arrive))
\end{lstlisting}

There are two possibilities: either the same customer visited
the deli twice, or two different customers came to the deli,
one for lunch and the other for dinner. We define an \textbf{Anglican} function
for each case. Note that this is the first time we switch from
Clojure to Anglican. The functions must be written in Anglican
(and hence defined using \texttt{defm} instead of \texttt{defn})
because they contain probabilistic forms \texttt{sample} and
\texttt{observe}.

\begin{lstlisting}[style=default]
(defm same-customer 
  "observe the same customer twice"
  [time-to-arrive-prior
   lunch-delay
   dinner-delay]
  (let [time-to-arrive
          (sample time-to-arrive-prior)]
    (observe (normal time-to-arrive time-sd) 
             lunch-delay)
    (observe (normal time-to-arrive time-sd)
             dinner-delay)
    [time-to-arrive]))

(defm different-customers
  "observe different customers"
  [time-to-arrive-prior
   lunch-delay
   dinner-delay]
  (let [time-to-arrive-1
          (sample time-to-arrive-prior)
        time-to-arrive-2
          (sample time-to-arrive-prior)]
    (observe (normal time-to-arrive-1 time-sd) 
             lunch-delay)
    (observe (normal time-to-arrive-2 time-sd)
             dinner-delay)
    [time-to-arrive-1 time-to-arrive-2]))
\end{lstlisting}

Both functions have the same structure: we first `guess' the
average arrival time and then observe the actual time from a
distribution parameterized by the guessed time. However, in
\texttt{same-customer} the average arrival time is the same for
both the lunch and the dinner, while in
\texttt{different-customers} two average arrival times are
guessed independently.

We are finally ready to define the query in \textbf{Anglican}.

\begin{lstlisting}[style=default]
(defquery deli [time-to-arrive-prior
                lunch-delay
                dinner-delay]
  (let [is-same-customer (sample (flip p-same))
        observe-customer (if is-same-customer
                            same-customer
                            different-customers)]
    {:same-customer is-same-customer,
     :times-to-arrive (observe-customer 
                        time-to-arrive-prior
                        lunch-delay
                        dinner-delay)}))
\end{lstlisting}

Performing inference on query \texttt{deli} computes the
probability that the same customer visited the deli twice, as
well as probability distributions of average arrival times for
both cases.

\subsection{Inference}
\label{seq:deli-inference}

Having defined the query, we are now ready to run the query
using an inference algorithm. Function \texttt{doquery} provided
by the \texttt{anglican.core} namespace accepts the inference
algorithm, the query, and optional parameters, and returns a
lazy sequence of samples. We use here the inference algorithm
called Lightweight Metropolis-Hastings (LMH). The algorithm is
somewhat slow to converge but can be used with any Anglican
query, and should be robust enough for our simple problem.

We bind the results of \texttt{doquery} to variable
\texttt{samples}, to analyse the results later. However, since
the sequence is lazy, no inference is performed and no samples
are generated until they are retrieved and processed.

\begin{lstlisting}[style=default]
(def samples (doquery :lmh deli nil))
\end{lstlisting}

To approximate the inferred distribution, we extract a finite
subsequence of samples. Many algorithms use an initial
subsequence of samples to converge to the target distribution.
Hence, we drop initial $N$ samples ($N$ is 5000 in the code
snippet below), and collect the next $N$ samples.

\begin{lstlisting}[style=default]
(def N 5000)
(def results (map get-result
                   (take N (drop N samples))))
\end{lstlisting}

Based on the collected samples we compute an approximation of
the posterior probability \texttt{p-same+} that the same
customer visited the deli twice.

\begin{lstlisting}[style=default]
(def p-same+ 
     (/ (count (filter :same-customer results))
        (double N)))
\end{lstlisting}
The \texttt{:same-customer} here represents a function
that looks up the entry with the same name in a given map.
The map stores the result of a single sampled execution,
and the \texttt{:same-customer} entry in the map
records whether the same customer visits the deli in the execution.
The \texttt{filter} function in \texttt{p-same+} 
goes through all the maps in \texttt{results}
and picks only the ones whose executions involve just one customer.

With the specified observations, \texttt{p-same+} is $\approx
0.12$. The probability is much lower than the
waiter's guess \texttt{p-same} ($\frac 2 3$).
Of course, the results may vary from run to run, and, for a given
algorithm, the accuracy depends on the number of samples we
decide to retrieve. 

\begin{figure}
    \includegraphics[trim={12pt 0 0 0},scale=0.7]{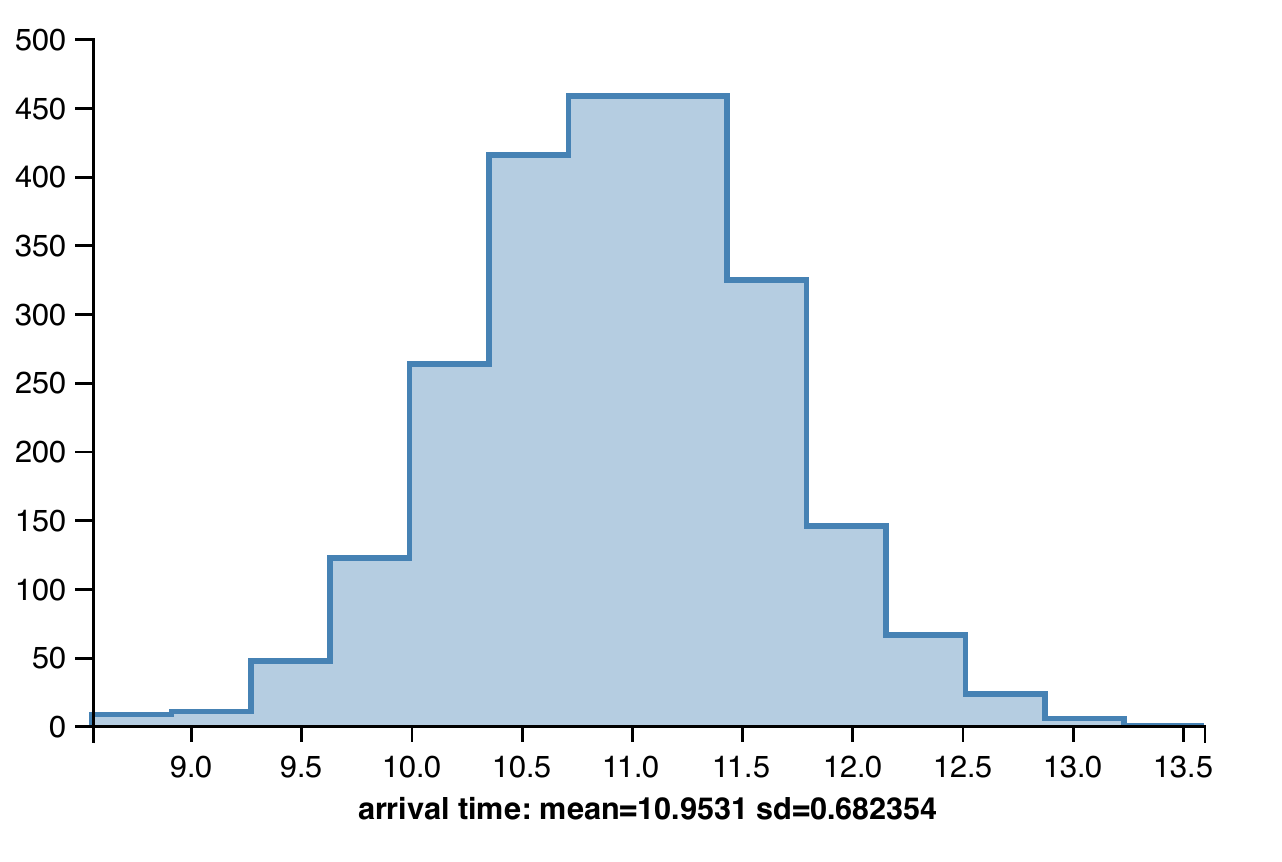}
    \caption{Arrival time distribution for a single customer.}
    \label{fig:dist-same}
\end{figure}

\begin{figure}
    \includegraphics[trim={12pt 0 0 0},scale=0.7]{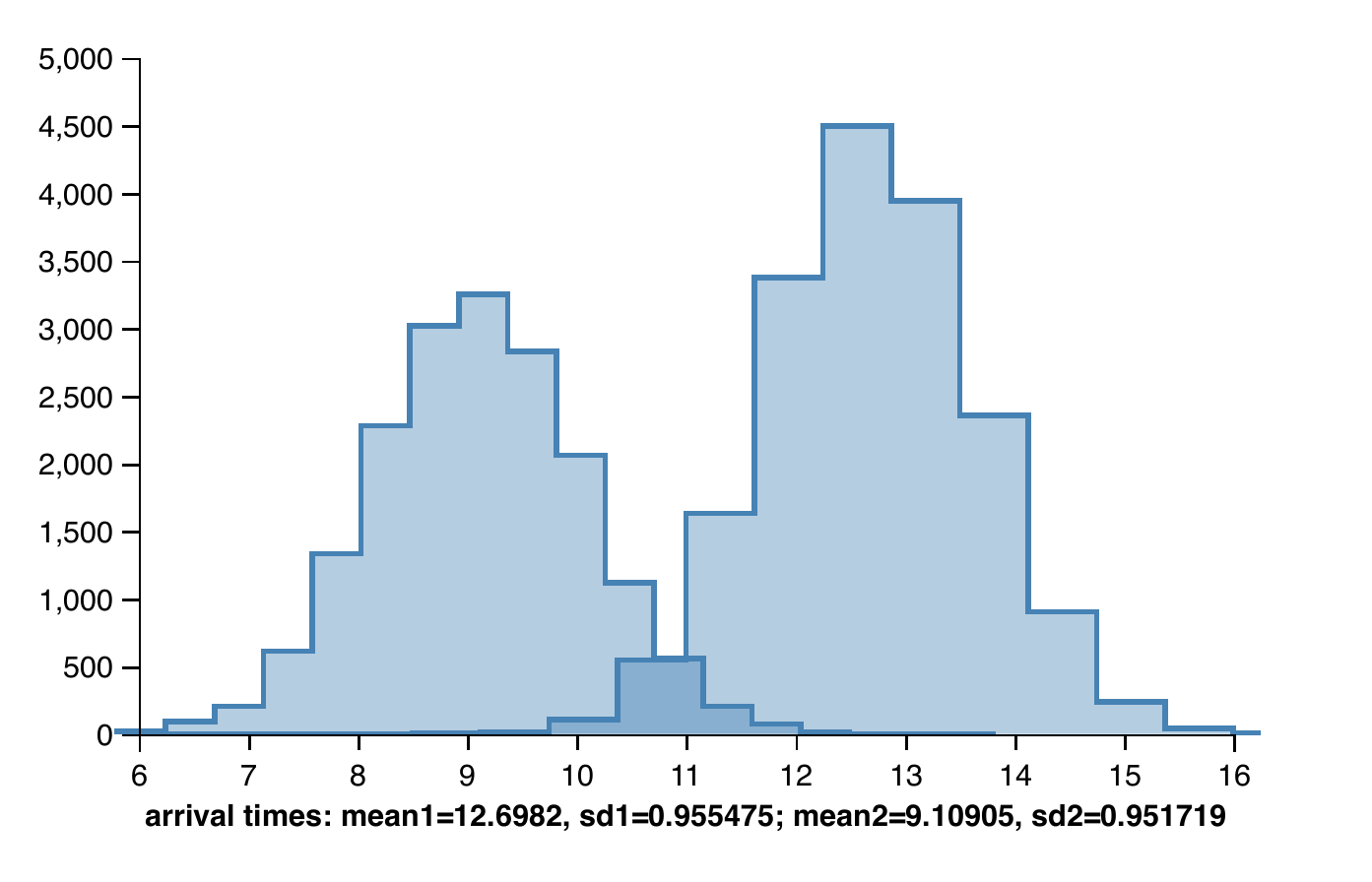}
    \caption{Arrival time distributions for two customers.}
    \label{fig:dist-diff}
\end{figure}

In addition to computing the posterior probability that the same
customer visited the deli twice, we may want to know the average
time (or times) to arrive. In Bayesian inference, it is common
to report distributions instead of `most likely' values. We use
query results in retrieved samples to approximate the
distributions, and plot distribution histograms for the same
customer visiting twice (Figure~\ref{fig:dist-same}) and
two different customers (Figure~\ref{fig:dist-diff})

\begin{lstlisting}[style=default]
;; single customer                      
(def time-to-arrive+ 
    (map (fn [x] (first (:times-to-arrive x)))
         (filter :same-customer results)))
(def mean-to-arrive+ (mean time-to-arrive+))
(def sd-to-arrive+ (std time-to-arrive+))

;; two customers
(def times-to-arrive+ 
     (map :times-to-arrive 
          (filter  
            (fn [x] (not (:same-customer x)))
            results)))
(def mean-1-to-arrive+
     (mean (map first times-to-arrive+)))
(def sd-1-to-arrive+
     (std (map first times-to-arrive+)))
(def mean-2-to-arrive+
     (mean (map second times-to-arrive+)))
(def sd-2-to-arrive+
     (std (map second times-to-arrive+)))
\end{lstlisting}

In addition to plotting the distribution histograms, we use
functions \texttt{mean}, \texttt{std} provided
in the \texttt{anglican.{\linebreak[0]}stat} namespace
along with other
useful statistical functions.
This is another illustration of advantage of tight
integration between Clojure and Anglican --- probabilistic
models are expressed in Anglican. However, processing of data and
results can rely on the full power of Clojure.

For inference, we chose to use  Lightweight
Metropolis-Hastings, perhaps somewhat voluntarily. A strength of
probabilistic programming is that models are separated from
inference. To switch to a different inference algorithm we just
need to pass a different value to \texttt{doquery}. For example,
we may decide to use Black-Box Variational Bayes (BBVB), which
may not work equally well for all probabilistic programs, but is
much faster to converge.

\begin{lstlisting}[style=default]
(def samples (doquery :bbvb deli nil))
\end{lstlisting}

We can still use samples and summary statistics with BBVB to
approximate the posterior distribution. However, variational
inference approximates the posterior by known distributions,
and we can directly retrieve the distribution parameters.

\begin{lstlisting}[style=default]
(clojure.pprint/pprint
  (anglican.bbvb/get-variational 
    (nth samples N)))
\end{lstlisting}

\begin{lstlisting}[style=default]
{S28209
 {(0 anglican.runtime.normal-distribution)
  {:mean 10.99753360180663,
   :sd 0.7290976433082352}},
 S28217
 {(0 anglican.runtime.normal-distribution)
  {:mean 12.668050292254202,
   :sd 0.9446695174790353}},
 S28215
 {(0 anglican.runtime.normal-distribution)
  {:mean 9.104132559955836,
   :sd 0.9479290526821788}},
 S28219
 {(0 anglican.runtime.flip-distribution)
  {:p 0.11671977016875924,
   :dist
   {:min 0.0,
    :max 1.0}}}}
\end{lstlisting}

We can guess that the variational distributions correspond to the
prior distributions in the \texttt{sample} forms, in the order
of appearance. However, it would help if we could use more
informative labels instead of automatically generated symbols
\texttt{S28209}, \texttt{S28217}, etc. Here the option to
specify identifiers for probabilistic forms explicitly comes
handy. If we modify the \texttt{sample} forms to use explicit identifiers
(only the forms are shown for brevity), the output becomes much
easier to analyse.

\begin{lstlisting}[style=default]
(defm same-customer 
      ...
      (sample :arrival-time-same
              time-to-arrive-prior)]
      ...)

(defm different-customers
      ... 
      (sample :arrival-time-first
              time-to-arrive-prior)
      ...
      (sample :arrival-time-second
              time-to-arrive-prior)]
      ...)

(defquery deli ...
      ...
      (sample :same-or-different
              (flip p-same))
      ...)
\end{lstlisting}

The output becomes much easier to interpret and analyse
programmatically.

\begin{lstlisting}[style=default]
{:arrivate-time-same
 {(0 anglican.runtime.normal-distribution)
  {:mean 10.99753360180663,
   :sd 0.7290976433082352}},
 :arrival-time-first
 {(0 anglican.runtime.normal-distribution)
  {:mean 12.668050292254202,
   :sd 0.9446695174790353}},
 :arrival-time-second
 {(0 anglican.runtime.normal-distribution)
  {:mean 9.104132559955836,
   :sd 0.9479290526821788}},
 :same-or-different
 {(0 anglican.runtime.flip-distribution)
  {:p 0.11671977016875924,
   :dist
   {:min 0.0,
    :max 1.0}}}}
\end{lstlisting}

This completes the case study, where we showed a probabilistic
programming solution of a problem, implemented and analysed in
Anglican and Clojure, using two different inference algorithms.

\section{Conclusion}
\label{seq:summary}

In this paper, we presented design and implementation internals
of the \emph{probabilistic programming system} Anglican. Implementing
a language is an interesting endeavour, in particular when the
language implements a new paradigm, in this case probabilistic
programming. Functional programming is a natural complement of
probabilistic programming --- the latter allows both concise and
expressive specification of probabilistic generative models and
efficient implementation of inference algorithms.

Implementing a probabilistic language on top of and in tight
integration with a functional language, Clojure, both helped
us to accomplish an ambitious goal in a short time span, and
provided important insights on structure and semantics of
probabilistic concepts incorporated in Anglican. Computational
efficiency and expressive power of Anglican owe to adherence to
the functional approach as much as to rich inference
opportunities of the Anglican environment.

\acks

This work was partially supported under DARPA PPAML through the
U.S. AFRL under Cooperative Agreement number FA8750- 14-2-0006,
Sub Award number 61160290-111668.

\bibliographystyle{abbrvnat}
\bibliography{refs}

\end{document}